# Improved iron-tolerance in recycled aluminum alloys via direct strip casting process

L. Jiang[1*], R.K.W. Marceau[1], T. Dorin[1]

[1]Institute for Frontier Materials, Deakin University, 75 Pigdons Road, Waurn Ponds, Victoria, Australia 3216

l.jiang@deakin.edu.au

**Graphical Abstract**

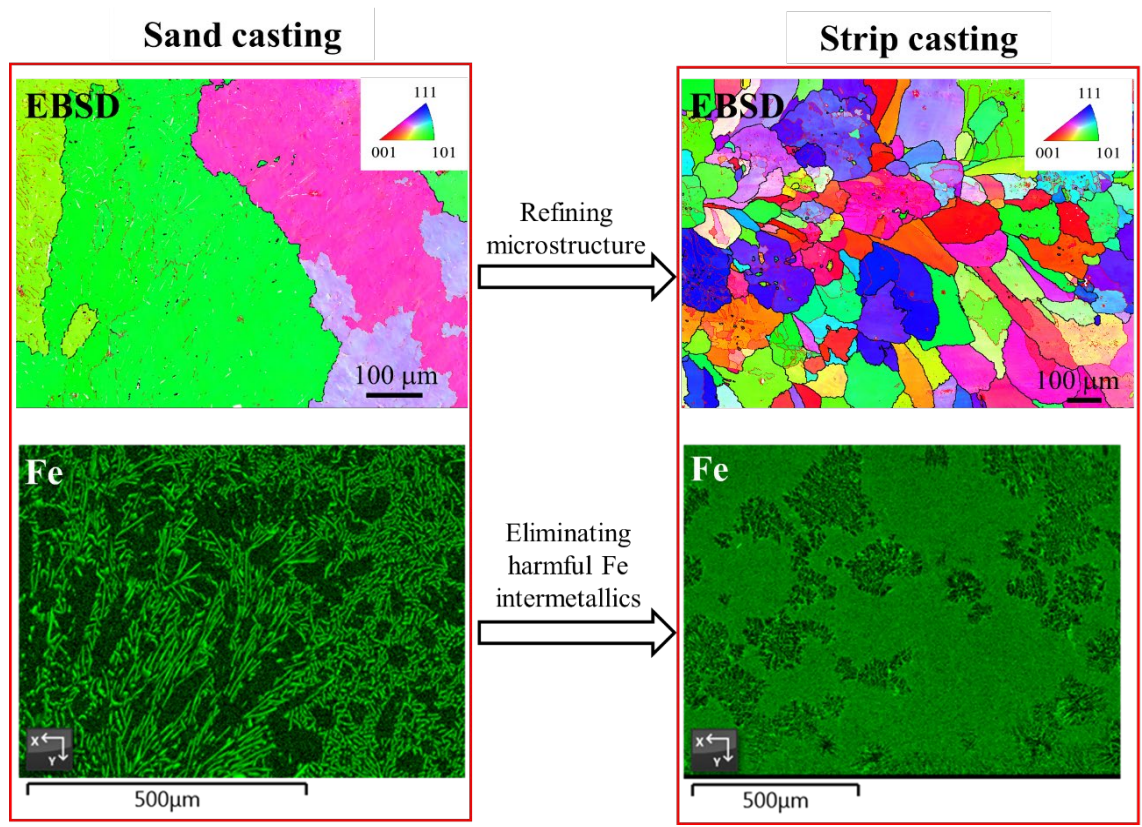

**Abstract**

Recycled aluminum alloys are pivotal for sustainable manufacturing, offering strength, durability, and environmental advantages. However, the presence of iron (Fe) impurities poses a major challenge, undermining their properties and recyclability. Conventional manufacturing processes result in coarse Fe-rich intermetallic compounds that limit the tolerance of Fe content and negatively influence performance of advanced aluminum alloys. To address this, rapid solidification techniques like direct strip casting have been explored. In this work, a detailed study of the strip cast microstructure was conducted by scanning electron microscopy, electron backscattered diffraction and atom probe tomography. Our results reveal that alloys produced by DSC exhibit significantly refined microstructures and are free from coarse Fe-rich intermetallics, thereby retaining the majority of Fe in solid solution. These findings indicate that strip casting significantly enhances Fe-tolerance in aluminum alloys, making it an attractive process for future aluminum recycling, with implications for sustainable high-performance applications.

## 1. Introduction

Recycled aluminum alloys are indispensable materials in today's sustainable manufacturing practices, offering a compelling combination of strength, durability, and environmental benefits. These alloys, derived from various sources of aluminum scrap, hold significant potential for reducing energy consumption and minimizing environmental impact when compared to primary aluminum production [1, 2]. However, their widespread adoption faces a critical challenge: the presence of iron (Fe) impurities, which tend to accumulate during recycling processes and severely compromise the alloys' properties and recyclability.

Fe, as the most common impurity in aluminum alloys, plays a notorious role in diminishing their corrosion resistance and mechanical properties. Unlike other metallic alloys, aluminum has a very limited solubility for Fe, maximum 0.05 wt.% at 650°C [3]. In conventional manufacturing, Fe mostly forms coarse intermetallic compounds, such as AlFe and AlFeMnSi, which not only exacerbate pitting corrosion but also limit the tolerance of Fe content in high-performance aluminum alloys. This constraint is particularly detrimental in aerospace and marine industries demanding low Fe levels, where the content is required to be less than 0.1 wt.% [2, 4]. Consequently, finding innovative solutions to enhance the Fe tolerance of recycled aluminum alloys is of paramount importance to promote their sustainability and utilization in critical applications.

One promising avenue for overcoming the challenges posed by Fe impurities in recycled aluminum alloys is through rapid solidification techniques that offer the advantage of refinement of the intermetallics. Direct strip casting (DSC) is an advanced near-net-shape casting technique that processes liquid aluminum directly into sheet, which offers significant cost reduction and energy savings [5]. The solidification rate of DSC is high, $10^2$ to $10^4$ °C/s [6], leading to the formation of fine-grain microstructure [5, 7]. In our previous work [8], direct strip casting has been proven to effectively refine Fe-rich intermetallic compounds in aluminum alloys and consequently enhance the alloy's corrosion resistance. This observation demonstrates the immense potential of direct strip casting in improving the recyclability of aluminum alloys. Nevertheless, a comprehensive examination of the microstructure generated through direct strip casting in aluminum alloys remains unexplored, which is essential for the future utilization of DSC within the aluminum recycling sector.

This contribution explores the effects of DSC on the microstructure, intermetallics, and solid solution composition of various Al-Fe alloys by using scanning electron microscopy (SEM), electron backscattered diffraction (EBSD), and atom probe tomography (APT). This work aims to pave the way for sustainable practices that not only extend the recyclability of aluminum but also enable the utilization of recycled alloys in high-performance applications where Fe content is traditionally a limiting factor.

## 2. Materials and experiments

In this study, we examined aluminum alloys with different Fe concentrations. The specific chemical compositions of these alloys are detailed in Table 1. Fe content of 0.1 wt.% was chosen because many high-performance aluminum alloys maintain a tolerance threshold below 0.1 wt.%. To further our research objectives, we intentionally elevated the Fe contents to 1.0 and 2.5 wt.%. This was done to assess the potential of direct strip casting in increasing Fe tolerance within aluminum alloys and to highlight the impact of Fe on both the microstructure and the overall properties of the material.

The experiments on direct strip casting were conducted using a lab-scale simulator, known as a dip tester, designed at Deakin University. This dip tester simulates the initial interaction between the molten material and the twin-roll caster's rolls during the twin-roll direct strip casting process [9]. The observed solidification rate reached approximately 500 °C/s. Each composition was produced through direct strip casting. Additionally, an Al-2.5Fe alloy was crafted using sand casting in a 3 kg rectangular mold, achieving a solidification rate of about 0.1 °C/s, similar to traditional industrial casting methods. The casting methods and cooling rate measurements are described in more detail in our previous work [8].

**Table 1.** Chemical compositions (wt.%) of the studied alloys.

| Alloy | Fe | Al |
|---|---|---|
| **Al-0.1Fe** | 0.1 | bal. |
| **Al-1.0Fe** | 1.0 | bal. |
| **Al-2.5Fe** | 2.5 | bal. |

Samples SEM and EBSD observations were prepared by taking sections perpendicular to the casting direction. The sectioned specimens were manually ground and polished with silicon carbide papers, followed by 6, 3 and 1 μm polycrystalline diamond suspensions. The initial polishing was conducted with active oxide polishing suspension (OPS) for 2 min. To improve the sample quality for the EBSD characterization, the samples were then polished by using vibration polishing with OPS for at least 12 hrs. SEM images were taken using the backscattered electron (BSE) detector in a JEOL JSM 7800F SEM instrument, equipped with energy dispersive spectroscopy (EDS) and EBSD detectors, with an operating voltage of 20 kV. For the EBSD experiments, the step sizes ranged from 0.5 μm. The EBSD scan data was analyzed using HKL Channel 5 software (Oxford Instruments HKL, Denmark).

APT experiments were conducted to determine the chemical composition of the solid solution as well as the local distribution of elements. These experiments were performed on a Local Electrode Atom Probe (LEAP 5000 XR, CAMECA) instrument with a pulse fraction of 20%, a pulse repetition rate of 250 kHz, a detection rate of 0.5%, and a specimen temperature of 25 K. APT samples were electro-polished with a standard two-step process [10]. APT data reconstruction and analysis was performed using CAMECA AP Suite 6 containing the Integrated Visualization Analysis Software (IVAS). Background subtraction was performed using the background correction tool in IVAS.

## 3. Results and discussion

Figure 1 shows the BSE images of the studied alloys, comparing those produced via strip casting (Fig. 1a-c) with the Al-2.5Fe alloy produced via sand casting (Fig. 1d). Notably, the alloys produced through strip casting exhibit an elongated grain structure aligned with the solidification direction. This is consistent with the previous literature [8, 11]. Interestingly, the strip cast microstructure is increasingly refined with increasing Fe content, as shown in Fig. 1 a-c. This phenomenon is attributed to the solute effect that is traditionally explained through growth restriction theory [12-14]. As solidification progresses, the Fe in the melt preferentially segregates in the liquid adjacent to the solid-liquid interface, which impedes the movement of the grain boundaries and thereby slows down the growth of the solid phase. Additionally, there are no discernible coarse particles in the Al-0.1Fe and Al-1.0Fe strip cast alloys, evident in Fig. 1 a & b, which suggests that most of the Fe in these alloys remains in solid solution, preventing it from diffusion and the formation of precipitates during the rapid solidification. Unlike the Al-0.1Fe and Al-1.0Fe strip cast alloys, the Al-2.5Fe alloy produced by direct strip casting exhibits a combination of eutectic and elongated structure. Furthermore, when examining the Al-2.5Fe alloy produced via sand casting (Fig. 1d), the formation of coarse intermetallics with a needle-like morphology in the inter-dendritic regions are observed. These intermetallics range in size from 20 to 300 μm with an area fraction of ~ 20%. In contrast, these coarse intermetallics are absent in the strip-cast Al-2.5Fe alloy, as shown in Fig. 1c.

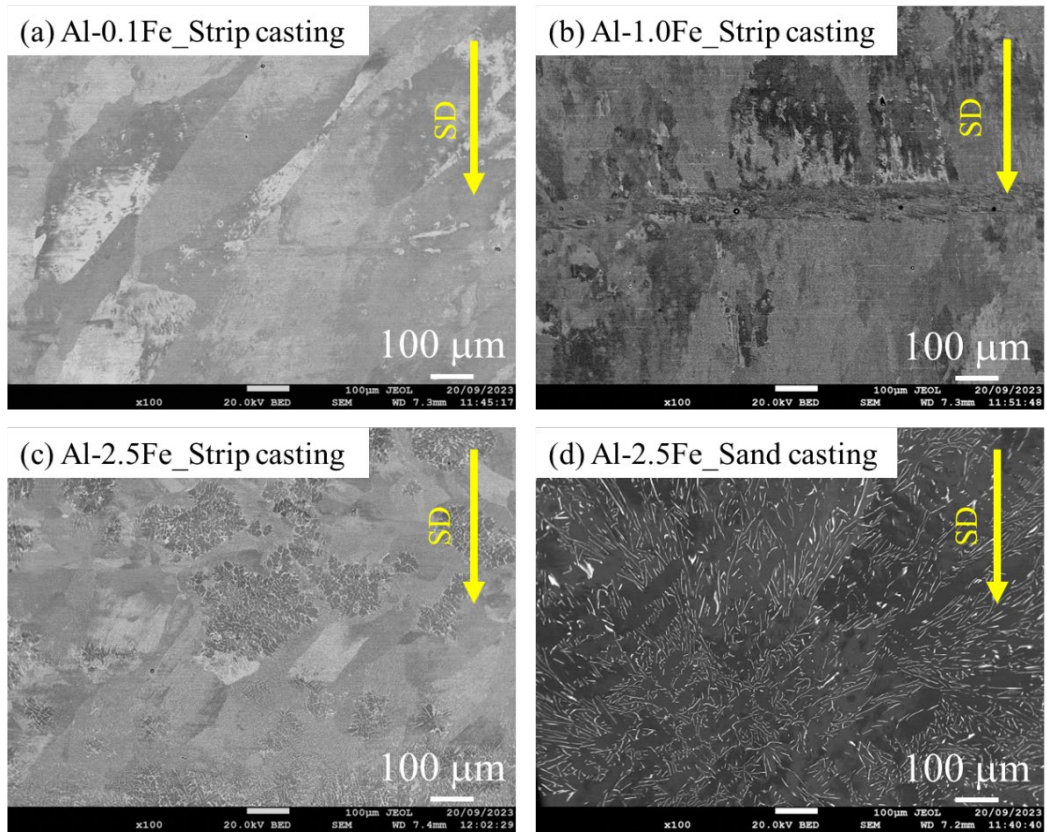

**Figure 1.** BSE imaging of the studied alloys: (a) Al-0.1Fe alloy produced by DSC, (b) Al-1.0Fe produced by DSC, (c) Al-2.5Fe produced by DSC, and (d) Al-2.5Fe alloy produced by sand casting. SD represents solidification direction.

To further investigate the difference in the microstructure of the Al-2.5Fe alloys produced by sand casting and strip casting, in-situ EBSD and EDS were carried out, as shown in Fig. 2. It can be seen from Fig. 2 a & b that the grains of the Al-2.5Fe alloy produced by sand casting are significantly coarser with an average diameter of 350 ± 121 μm, which is about 3 times larger than that observed in the Al-2.5 alloy produced by strip casting (111.7 ± 54.5 μm). Grain refinement through rapid solidification associated with strip casting has been reported previously in the literature [5, 15], where high undercooling is achieved.

The EDS mapping result in Fig. 2c suggests that the coarse needle-shape intermetallic phases in the sand cast Al-2.5Fe alloy that are enriched in Fe are likely the $Al_3Fe$ (sometimes described as $Al_{13}Fe_4$) phase that is commonly seen in Al-Fe binary alloys [8, 16, 17]. Due to the slow solidification of sand casting, the following high-temperature equilibrium eutectic reaction occurs: $Liquid \rightarrow \alpha_{Al} + Al_3Fe$, which is known to occur over the temperature range from 652 – 655°C in aluminum-rich alloys [16]. However, these coarse Fe-rich particles are not observed in the alloy produced by direct strip casting (Fig. 2d). Instead, the EDS result in Fig. 2d shows that there are refined eutectic Fe-rich phases in the Al-2.5Fe alloy produced by strip casting. Upon closer examination of Fig. 2 b & d, it can be found that the Fe-rich phase primarily forms along the low angle boundaries, possibly dendritic boundaries. According to literature, the refined Fe-rich phase could be identified as the metastable compound $Al_9Fe_2$, previously observed in rapidly solidified Al-Fe alloys, especially at cooling rates exceeding 20°C/s [16]. The emergence of the metastable Fe phase during rapid solidification can be attributed to the progressively increasing supercooling and to the change in conditions for nucleation and growth [16].

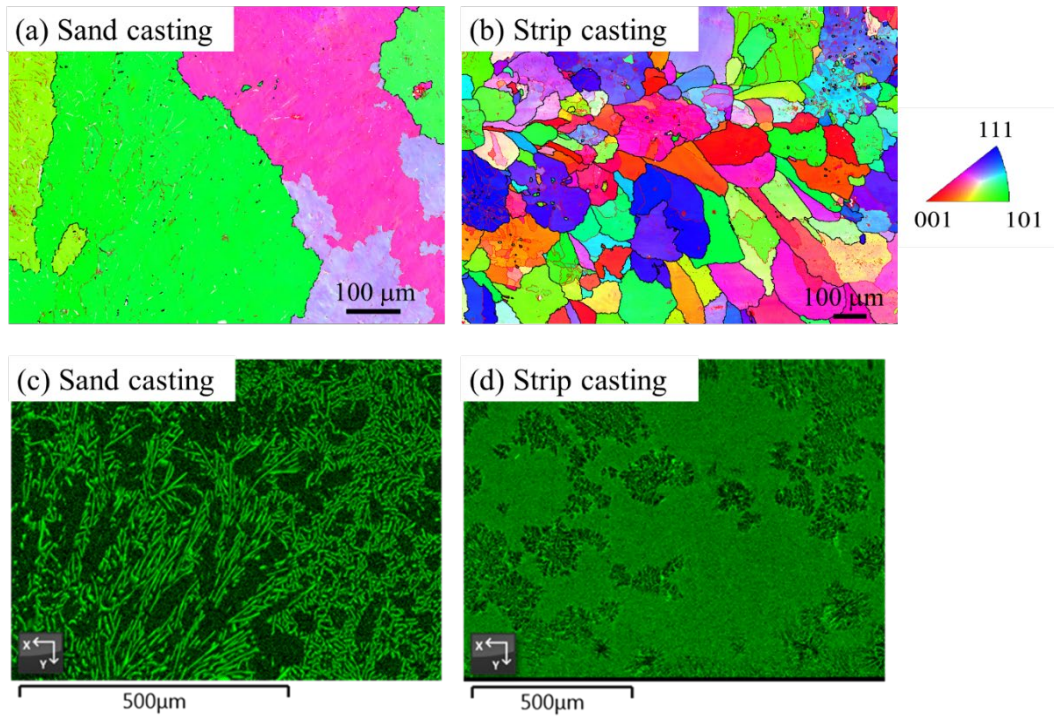

**Figure 2.** (a) & (b) are the EBSD-derived grain orientation maps for the Al-2.5Fe alloys produced by sand casting and strip casting, respectively. The inverse pole figure (IPF) color legend shown on the right applies to both EBSD images. (c) and (d) are the corresponding EDS maps for the Fe distribution within the same regions shown in (a) and (b), although note the scale bar changes.

Atom probe tomography (APT) is the most accurate tool for near atomic resolution concentration measurement [18]. To determine the Fe content in solid solution, APT analyses were carried out to evaluate the matrix Fe concentration of the Al-2.5Fe alloy material produced by sand casting and strip casting. Fig. 3 shows three-dimensional reconstructions that map the atomic-scale distributions of Al and Fe atoms (in cyan and pink colors, respectively) for these materials. Notably, both Al and Fe are uniformly dispersed within these data volumes for both material conditions. It is not surprising that no discernible presence of Fe-rich compounds was captured given the analysis volume dimensions are on the order of a few tens of nm as shown in Fig. 2. This suggests that both APT data represent the matrix of the two alloys, proving invaluable for determining the chemical composition of the solid solution. In APT analysis, time-of-flight-based mass-to-charge state spectrum peak overlap occurs for $AlH^+$ and $Fe^{2+}$ isotopes at 28 Da, and for $AlH^+$, $AlH^{2+}$ and $Fe^{2+}$, at 29 Da. Here, this issue was carefully addressed using the peak decomposition tool within IVAS, by comparing the relative natural abundance of the elemental isotopes. Table 2 gives the decomposed and background-corrected composition of the matrix of the Al-2.5Fe alloys produced via sand casting and strip casting. It can be seen that the amount of Fe in solid solution of the strip cast Al-2.5Fe alloy is ~ 0.90 ± 0.06 at.% (1.90 ± 0.13 wt.%), which is significantly higher than that of the alloy produced by sand casting at 0.21 ± 0.01 at.% (0.44 ± 0.02 wt.%). This indicates that more Fe is retained in the solid solution during strip casting compared to traditional sand casting. Conversely, the loss of Fe (~ 0.60 wt.%) in the solid solution of the Al-2.5Fe alloy produced by strip casting still results from the formation of metastable $Al_9Fe_2$ compounds, as shown in Fig. 2d. Nevertheless, comparison of the SEM and APT results for the two alloy manufacturing processes highlights the capability of direct strip casting to avert the formation of coarse Fe intermetallics by forcing Fe into the solid solution by almost 10 times as much. The measured Fe content in solid solution in the sand-cast alloy exceeded the equilibrium solubility (0.03 at.%) [19, 20] and this could be due to a couple of factors. Firstly, non-equilibrium solidification could occur in some regions where Fe atoms might get trapped within the solid solution [16]. Secondly, there could be Fe micro-segregation commonly seen in traditional casting processes [21, 22], particularly inter-dendritic areas where the Fe concentration is higher. Notwithstanding, the marked difference observed between the two samples clearly indicates that strip casting retains a higher amount of Fe in solid solution.

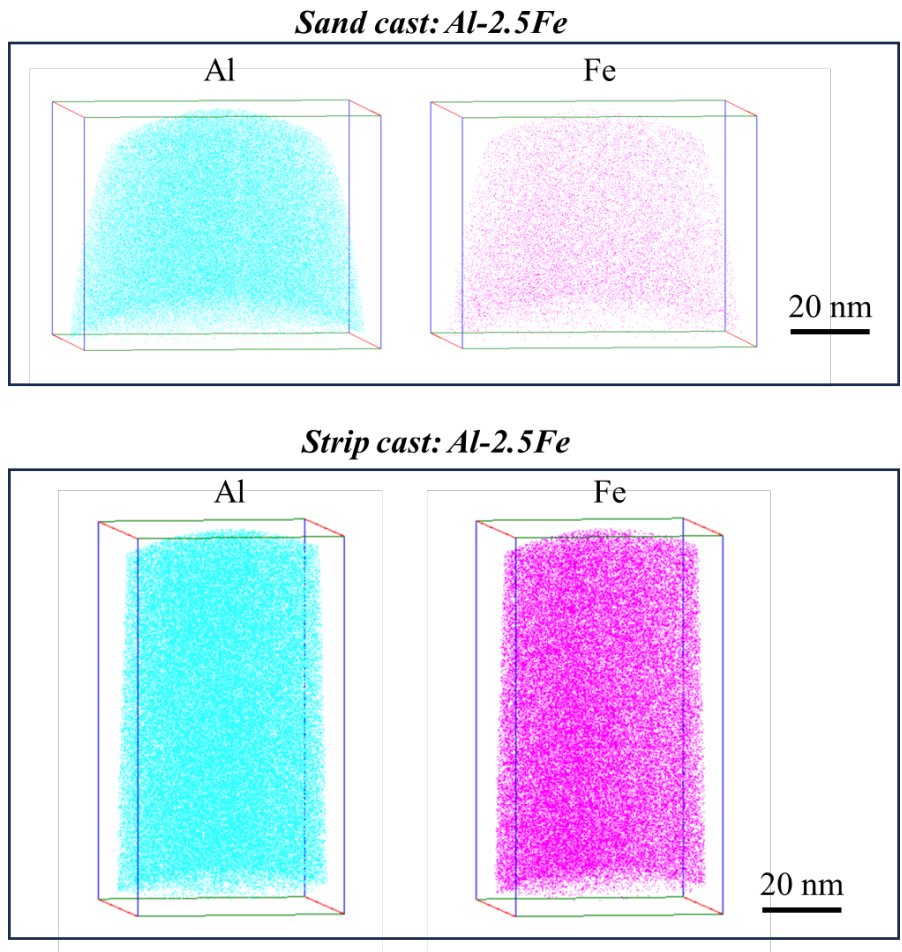

**Figure 3.** Reconstructed APT maps of Al (cyan) and Fe (pink) for the Al-2.5Fe alloys produced by (a) sand casting and (b) strip casting.

**Table 2.** Summary of the background-corrected chemical composition (at.%) of the matrix of the Al-2.5Fe alloy produced by sand casting and strip casting, derived from the APT results.

| Casting | Al | Fe | Other elements |
| --- | --- | --- | --- |
| **Sand casting** | 97.42 ± 0.15 | 0.21 ± 0.01 | Bal. |
| **Strip casting** | 95.86 ± 0.18 | 0.90 ± 0.06 | Bal. |

In summary, the detailed microstructural analyses presented in this work demonstrates that direct strip casting has the capability to refine the grain microstructure, free from coarse intermetallics, and retain a considerable amount of Fe in solid solution in Al-Fe alloys. This contrasts with the material from traditional sand casting, where coarse and needle-shaped Fe intermetallics dominate the microstructure with much less Fe in the matrix. The absence of these coarse and needle-shaped Fe intermetallics in the strip-cast aluminum alloys has been reported to improve the alloy's corrosion resistance significantly [8]. In addition, as the Fe content in solid solution increases, the strip-cast microstructure is increasingly refined, which further improves the corrosion performance of the alloy [23, 24]. Furthermore, increasing the Fe content in solid solution coupled with grain refinement can enhance the alloy strength through a combination of solid solution hardening and Hall-Petch strengthening, and the absence of coarse and needle-like Fe particles can serve to avoid ductility loss. These properties are especially advantageous for alloys with high Fe content that demanding high ductility, such as 1xxx aluminum alloys [16]. As such, this work demonstrates the transformative potential of direct strip casting in aluminum alloy recycling, particularly to enhance the tolerance of Fe typically considered a harmful impurity, yet frequently accumulated during aluminum recycling processes.

## 4. Conclusions

1) Alloys produced by strip casting exhibited an elongated grain structure aligned with the solidification direction, with the microstructure increasingly refined with higher Fe content.
2) In contrast, the sand-cast Al-2.5Fe alloy exhibited coarser grains with an average diameter approximately three times larger than strip-cast Al-2.5Fe. Additionally, coarse needle-shape Fe-rich intermetallic phases were found in the sand cast alloy, while these were notably absent in the strip-cast counterpart.
3) APT analysis of the strip-cast Al-2.5Fe alloy measured the Fe content in solid solution to be $1.90 \pm 0.13$ wt.%, which is significantly higher than that measured in the sand-cast alloy ($0.44 \pm 0.02$ wt.% Fe).
4) Rapid solidification during direct strip casting can substantially enhance the tolerance of aluminum alloys to Fe, rendering it an appealing process for future aluminum recycling endeavors.

## 5. Acknowledgements

The authors acknowledge the use of instruments at the Deakin Advanced Characterisation Facility. Funding from the Australian Research Council Discovery Project grant scheme (DP130101887) is also acknowledged.

Conflict of Interest Statement: On behalf of all authors, the corresponding author states that there is no conflict of interest.